# Design and Implementation of a Hybrid Wireless Power and Communication System for Medical Implants

Ali Khaleghi
Department of electronic systems (IES)
*Norwegian University of Science and Technology*
Trondheim, Norway
ali.khaleghi@ntnu.no

Aminolah Hasanvand
Department of electronic systems (IES)
*Norwegian University of Science and Technology*
Trondheim, Norway
aminolah.hasanvand@ntnu.no

Ilangko Balasingham
Department of electronic systems (IES)
*Norwegian University of Science and Technology*
Trondheim, Norway
ilangko.balasingham@ntnu.no

*Abstract*— Data collection and analysis from multiple implant nodes in humans can provide targeted medicine and treatment strategies that can prevent many chronic diseases. This data can be collected for a long time and processed using artificial intelligence (AI) techniques in a medical network for early detection and prevention of diseases. Additionally, machine learning (ML) algorithms can be applied for the analysis of big data for health monitoring of the population. Wireless powering, sensing, and communication are essential parts of future wireless implants that aim to achieve the aforementioned goals. In this paper, we present the technical development of a wireless implant that is powered by radio frequency (RF) at 401 MHz, with the sensor data being communicated to an on-body reader. The implant communication is based on two simultaneous wireless links: RF backscatter for implant-to-on-body communication and a galvanic link for intra-body implant-to-implant connectivity. It is demonstrated that RF powering, using the proposed compact antennas, can provide an efficient and integrable system for powering up to an 8 cm depth inside body tissues. Furthermore, the same antennas are utilized for backscatter and galvanic communication.

## I. INTRODUCTION

By developing biomedical sensors and adding communication and powering techniques, implant devices can be used for real-time measurement of sensory data to monitor health status and enable prevention and early treatment. Approved sensors for human healthcare include heart pacemakers, ECG and heart accelerometer sensors, glucose sensors for diabetics, and brain signal sensors for amputees and brain-machine interface devices. Some require high data rates sensing and transmission, such as brain-implanted electrodes and endoscopy camera capsules, while others require moderate or low rate sequential sensing. Most devices are battery-powered and offer wireless charging. Implants can be classified as superficial (under the skin or <2cm) or deep (>10cm). Efficient approaches to transferring power to deep implants are needed for monitoring organ health in transplant patients. Fig.1 shows a conceptual drawing of a patient with multiple implants and different connectivity links, such as conductive and radio techniques. In this example, the implant network consists of multiple Implant Wireless Local Area Networks (IWLAN) that collect local sensory data and communicate among each other to validate the data authenticity, while a central implant hub collects the multiple implants' data and communicates the data to an on-body wearable hub. To guarantee the system operation and longevity, that is essential for permanent implants, it is required to wirelessly charge the implant and use extremely low-power communication approaches for efficient connectivity. Communication consumes significant power and space in an implant device that is addressed in this paper using our innovative passive RF backscatter and galvanic impulse [1]–[3]. The backscatter technique involves a communication approach where an external remote reader device applies the power, and the remote tag data is transmitted by the reflection of the signal from the implant, which uses a simple diode or switch device to alter the implant's antenna reflection impedance [4]. A high data rate wireless RF backscatter method for transmitting image or video data from an endoscope capsule has been proposed and validated, which can handle data rates of 10-30 Mbps without using any active transmitter in the implant [5]. Another low-power approach proposed by the authors relies on using galvanic or conduction methods [1], [6], where the transmitted signal forms a baseband impulse generated by discharging a capacitor in the conductive medium of the body. In this approach, the power consumption for each data bit is reduced to at least 1.4 nJ/bit. The method is appropriate for low-rate communication with an active battery or wirelessly charged implant capacitor energy source. The system has been tested for communication among pacemaker implants in the cardiac chambers, where intra-body communication has been demonstrated. In this paper, a simultaneous communication and wireless powering technique for an implant has been developed, where the RF power is transferred for deep implant powering, RF backscatter is used for communication with the target implant. The same implant (s) also communicates data using active ultra-low power impulse techniques based on the galvanic to the other local implants via the galvanic impulse link. The design, simulation, and test process are presented as follows.

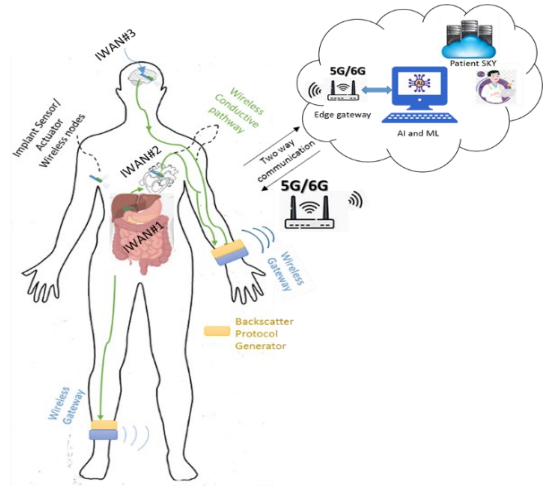

Fig. 1. Illustration of an implant communication network with on-body reader and wireless loacl implant network ues galvanic communication

## II. SIMULATION MODEL

The full-wave electromagnetic simulations are conducted using FDTD by implementing the homogeneous multilayer skin, fat, and muscle tissues. Precise modeling is employed within the 100-700 MHz frequency range to identify the optimal frequency, implant size, and depth for achieving maximum power coupling. The implanted antenna features an electrode configuration consisting of two circular electrodes with a diameter of 7 mm, a thickness of 1 mm, and a coupling gap of 0.1 mm [7]. The separation distance between the two electrodes varies (5, 10, 20, 30, 40 mm) to account for different implant lengths. One notable advantage of using electrodes is their ability to extend the virtual size of the antenna beyond its physical dimensions by utilizing medium conductivity. Additionally, electrode-based antennas are easily integrable with metal implants and conform to the geometry and size of the implant [7].

On the other hand, the on-body antenna comprises a pair of two square patches with a size of 15×15 mm². These patches are fed with a transmission line from the center, where they are coupled to the biological tissues with a gap of 1 mm [8]. The distance between these patches varies within the range of 30, 50, 70, and 100 mm, accounting for the size of the external antenna.

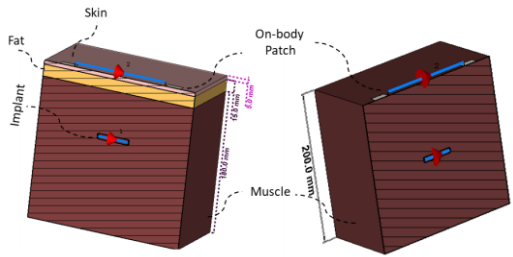

Fig. 2. Simulation model with a multilayer (skin, fat, muscle), implant antenna and on-body antenna are illustrated

Fig.2 shows the simulation models with the implant electrode antenna and the on-body patch antenna in multilayer tissues. The calculations are parametrically set to provide a comprehensive insight into the multiple effects of implant size, on-body patch size, distance, also antenna impedances. Figure 3-a shows the capsule antenna impedance ($R+jX$), in sample depth of 100 mm, for different capsule sizes simulated for the frequency (100-700 MHz). For instance, the capsule with a length of 20 mm, has resonance at 401 MHz with R=35Ω. Fig. 3-b shows the on-body antenna impedance for different lengths. The coupling between the on-body and implant electrode for the matched antennas is shown in Fig. 4 at a frequency of 401 MHz (MICS band) at 10 cm depth for the different implant and on-body antenna lengths. Additionally, the coupling for an implant length of 20 mm and on-body antenna size 100 mm for different frequencies 100-600 MHz are also illustrated, where at the range of 400 MHz to 600 MHz the coupling link is significantly higher than the lower frequencies at 100-300 MHz.

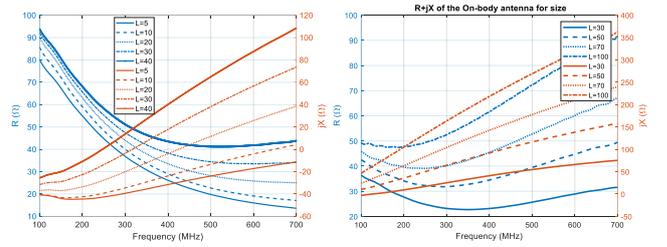

Fig. 3. a) Calculated implant antenna impedance in muscle tissue for different implant lengths, L=5, 10, 20, 30, 40 mm. b) Impedance of on-body patch versus frequency for different patch lengths.

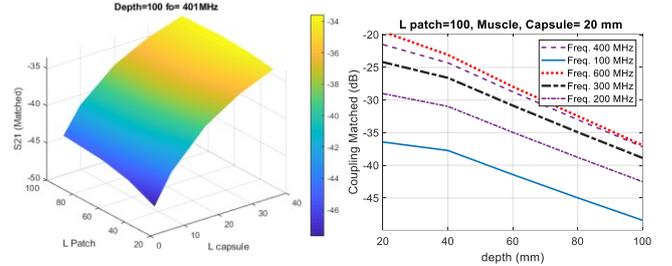

Fig. 4. Coupling between on-body antenna and the implant for matched scenario versus size of implant and on-body b) coupling with depth for different frequencies (impalnt size 20mm and on-body 100 mm).

## III. COMMUNICATION USING BACKSCATTER AND GALVANIC LINK

The sample designed implant system with three electrodes is shown in Fig. 5, the implant uses one common ground electrode for wireless powering, wireless backscatter, and galvanic link. A second electrode is used for simultaneous wireless powering and backscatter. A voltage doubler rectifier diode (HSMS 285C) with an optimized matching circuit at the 400 MHz band converts RF energy to DC and stores energy in a 330 pF capacitor. The stored energy runs an ultra-low power MEMS oscillator (SiT1569, with a minimum voltage of 1.8V and Power 3 microwatt) and an RF switch (ADG 902, 1.65 V, 165 nanowatts). The total power consumption is only 3 microwatts, and the external RF power continuously charges the capacitor. As the power requirement is minor and the goal is to keep the electronics simple, no power management unit (PMU) or voltage booster/regulator is used. Thus, the circuit starts operating once the received voltage passes the threshold of 1.8 V. A sensor data source activates the MEMS oscillator to modulate the MEMS oscillations that control the RF switch for backscatter and creates a subcarrier frequency. The applied RF energy is harvested in one data cycle and reflected in the other to activate the backscatter. In the process of switching, a portion of the energy charged in the capacitors can be released into the biological medium using the third conductive electrode to communicate the implant data to the surrounding implants for intra-body communication or synchronization in the local implant network, up to distances of 5 cm from the implant with a threshold detector scheme. Therefore, the other implants can receive information from the main hub implant, providing a link for implant-to-implant connectivity. Further efforts can be used to increase the connectivity range and add multiple access techniques to address the implants.

The measurement of the implant is conducted in phantom experiments mimicking biological muscle tissue at 400 MHz.

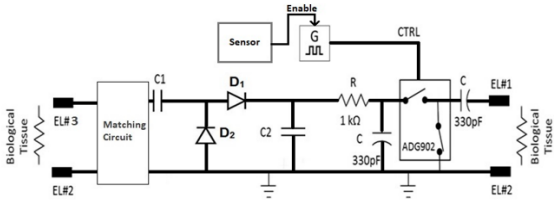

Fig. 5. Designed implant electronic circuit for wireless power, backscatter communication and galvanic connectivity.

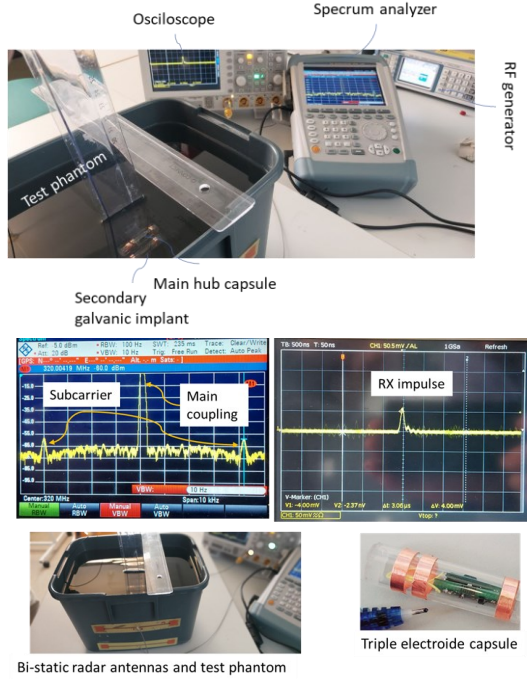

Fig. 6. a) measurement setup including signal generator at 401 Mhz, spectrum analyzer as receiver and text phantom b) received bckscatter spectrum at 8.5 cm, received galvanic impulse at 5 cm from the main capsule, reader antennas on the phantom and fabricated capsule mockup.

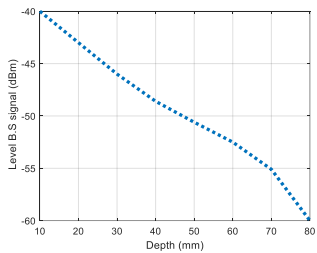

Fig. 7. Measure backscatter signal level versus distance for the transmitter power of 23 dBm in a phantom medium mimicking muscle tissues at 401 MHz.

The on-body reader transmits a 200 mW continuous wave signal at 401 MHz, and the calculated received power is -10 dBm (100 microwatts) at a depth of 10 cm. The full-wave rectifier diode of HSMS 285C can provide 40% of the received power to DC power (thus, 40 microwatt DC power is available), in which a minimum voltage of 1.8 V is required to run the low-power onboard electronics, achieved with the matching circuit of Q (min)=10. This means the absorbed power must provide sufficient voltage to drive the electronics without implementing a PMU. The ultra-low power nature of the communication, which is 1.4 pJ/bit for the galvanic link and 165 nW for switch operation and backscatter, is sufficient to run a sensor with power needs up to 35 microwatts at a distance of 10 cm without an additional PMU. Figure 6 shows the measurement setup, including a transmitter antenna (a patch-loaded dipole on the phantom surface with a gap of 2 mm to the liquid phantom), a second similar antenna as the receiver antenna in the bi-static radar reader setup with a distance of 5 cm, which reduces the direct coupling between on-body antennas to -25 dB. Additionally, a second implant is developed with two metal electrodes in direct contact with the phantom to obtain the impulses produced by the main implant. The received signal by the second implant is recorded using a digital sampling oscilloscope by implementing a low-pass filter to eliminate the RF coupling in the DSO. The developed implant is also illustrated. The measurements are conducted for different distances of the implant from the on-body transceiver antennas. The level of the subcarrier signal in the backscatter link is measured, as shown in Figure 7, attesting to the wireless powering of the implant's onboard electronics for depths up to 8 cm. The backscatter link loss is a linear decay of about 2.9 dB/cm.

## IV. Conclusion

A wireless implant powered by radio frequency (RF) at 401 MHz can communicate data using two simultaneous wireless links, the RF backscatter for implant-to-on-body communication and the galvanic link for intrabody implant-to-implant connectivity. The proposed system provides efficient and integrable wireless powering, sensing, and communication for up to 8.5 cm depth inside the body, utilizing compact antennas for backscatter and galvanic communication. The system has the potential to collect and process sensory data for early detection and prevention of chronic diseases without implant battery and assist in multi-node implant connectivity. The system's main feature is its simple implant circuitry and near-zero power connectivity.

## Acknowledgment

The work has been supported by the project Brain-Connected inteRfAce TO machineS (B-CRATOS), (https://www.b-cratos.eu) under grant 965044, Horizon 2020 FET-OPEN.